\documentclass[12pt]{article}
\usepackage{graphicx}
\usepackage{amsmath}
\usepackage{amsfonts}
\newcommand{\vect}[1]{ \mathbf{#1} }
\newcommand{\veps}{ { \varepsilon } }
\newcommand{\diff}{ \mathrm{d} }
\newcommand{\equal}{\stackrel{\mathrm{def}}{=}}
\newcommand{\ALD}{Abraham--Lorentz--Dirac\ }
\makeatletter
    \renewcommand*{\@fnsymbol}[1]{\ensuremath{\ifcase#1\or
 \textrm{a}\or
        \textrm{b}\or \textrm{c}\or \mathsection\or \mathparagraph\or
 \|\or
        **\or \dagger\dagger  \or \ddagger\ddagger \else\@ctrerr\fi}}
\makeatother

\title{Nonuniqueness of nonrunaway solutions of \ALD equation in an
  external laser pulse}

\author{A.~Carati\thanks{Universit\`a di Milano, Dipartimento di Matematica, 
    Via Saldini 50, 20133 Milano, Italy} \and
  M.~Stroppi\thanks{Universit\`a di Milano, Corso di laurea in Fisica,
    Via Celoria 12, 20133 Milano, Italy} }

\date{\today}
  
\begin{document}

\maketitle

\begin{abstract}
  In the paper \cite{carati95} it was shown that, for motions on a
  line under the action of a potential barrier, the third-order \ALD
  equation presents the phenomenon of nonuniqueness of nonrunaway
  solutions. Namely, at least for a sufficiently steep barrier, the
  physical solutions of the equation are not determined by the
  ``mechanical state'' of position and velocity, and knowledge of the
  initial acceleration too is required. Due to recent experiments,
  both in course and planned, on the interactions between strong laser
  pulses and ultra relativistic electrons, it becomes interesting to
  establish whether such a nonuniqueness phenomenon extends to the
  latter case, and for which ranges of the parameters. In the present
  work we will consider just the simplest model, i.e., the case of an
  electromagnetic plane wave, and moreover for the \ALD equation dealt
  with in the nonrelativistic approximation.  The result we found is
  that the nonuniqueness phenomenon occurs if, at a given frequency of
  the incoming wave, the field intensity is sufficiently large. An
  analytic estimate of such a threshold is also given. At the moment
  it is unclear whether such a phenomenon applies also in the full
  relativistic case, which is the one of physical interest.
\end{abstract}

%%\begin{keyword}
\textbf{Keyword}: \ALD equation, non uniqueness, electron scattering

%PACS:    05.20.--y: 47.27.-Eq  
%%\end{keyword}

\section{Introduction}\label{sec:intro} 

Abraham in ref.~\cite{abraham} and Lorentz in ref.~\cite{lorentz} (for
the later Dirac relativistic version see ref.~\cite{dirac38}),
proposed the following equation in order to describe the motion of a
radiating electron
\begin{equation}\label{eq:ald}
  m\ddot {\vect x} = \vect F(\vect x,\dot{\vect x} ) +
  \frac{2e^2}{3c^3} \dddot{\vect x} \ ,
\end{equation}
where $m$, $e$ and $c$ are the mass of the electron, its charge and
the speed of light respectively, while $\vect F(\vect x,\dot{\vect
  x})$ is the force acting on the electron, for example the Lorentz
force due to an incoming electromagnetic wave.  As is well known, the
solution of the equation for generic initial data diverge as $e^{\frac
  t\veps}$ for $t\to+\infty$: so they are physically absurd, since
they keep continuing to accelerating also if the force (i.e., the
electromagnetic pulse) vanishes. This inclined theorists to discharge
such an equation, replacing it by some suitable approximationt, for
which runaway solutions don't show up. The so called Landau--Lifschitz
approximation (see ref.~\cite{landaulifschitz}, or the more recent
ref.~\cite{dipiazzaLL}) was very recently tested in some experiments
of interaction between a beam of ultra relativistic electrons with
strong laser pulses. The agreement between theoretical prediction and
experimental data was not completely satisfactory (see
ref.~\cite{cole,dipiazza}).

One might think that the use of the original equation (\ref{eq:ald})
could give better agreement. The proposal advanced by Dirac in order
to overcome the problem of the runaway solutions, was to admit that
the only physically significant solutions are the ones for which the
accelerations $\ddot{\vect x}$ tends to zero as $t\to+\infty$.  From
the mathematical point of view, one has then to deal no more with a
Cauchy problem, for which existence and uniqueness of solutions are
granted, but with a boundary value problem, in which are given the
mechanical data of position and velocity at $-\infty$, and the
acceleration at $+\infty$.  For boundary problems uniqueness is not
granted, i.e., having fixed the mechanical data before the
interaction, there might exist several solutions which satisfy the
nonrunaway condition $\ddot{\vect x}\to 0$. Any approximation, as the
Landau--Lifschitz one, which instead admit just one solution, will be
a poor one in that regime.

So, it is of importance to understand whether, and eventually in what
regime, the nonuniqueness of the nonrunaway solutions shows up for the
\ALD equation. In this paper we investigate such a problem for the
nonrelativistic version (\ref{eq:ald}) of the \ALD equation, in the
case of an incoming electromagnetic plane wave. We will show through
numerical computations that there exists a threshold in the intensity
of the field, above which nonuniqueness occurs. Some numerical checks
are also performed, to control whether below threshold the
Landau--Lifschitz approximation is sound. It appears that also well
below the threshold the two equations lead to very different
behaviors, with the electron radiating, according to the Dirac
equation, much less energy than according the Landau--Lifschitz
approximation.

The paper is organized as follows. In Section~\ref{sez:model} we
describe the model studied, while in Section~\ref{sez:nonuni} we give
an analytic estimate of the region of parameters in which 
nonuniqueness is expected to occur. In Section~\ref{sez:numeric} we
illustrate the numerical results, and in Section~\ref{sez:landau} a
comparison with the Landau-Lifschitz approximation is
given. The conclusions follow.

\section{The model}\label{sez:model}

We consider the case of the interaction of an electron, described by
the \ALD equation, with an external electromagnetic linearly polarized
plane wave.  We will take the $x$ axis as the direction of the wave
propagation, the $y$ axis as the direction of electric field and
finally the $z$ axis as the direction of the magnetic field. In the
Coulomb gauge, the scalar potential vanishes, while the vector
potential $\vect A$ takes the form $ \vect {A} = (0, F(x-ct), 0)$,
being $ F (x-ct) $ an arbitrary function, and $c$ the speed of
light. To be definit, we model the electromagnetic pulse by choosing
$F(\xi)=A\exp(-\xi^2/2\sigma)\cos(k\xi)$, although every choice with $F$
vanishing  sufficiently fast at infinity would give the same qualitative
results. So, the electromagnetic field takes the form
\begin{equation*}
  \left\{
  \begin{aligned}
      \vect{E}(\vect{r},t)& = -F'(x-ct) \, \hat{\vect e_y}
      \\ \vect{B}(\vect{r},t) &= \quad F'(x-ct) \, \hat{\vect e_z} \ ,
  \end{aligned} 
  \right.
\end{equation*}
where $\hat{\vect e_y}$ and $\hat{\vect e_z}$ are  unit vectors
directed as the $y$ and $z$ axis respectively, while $F'$ denotes the
derivativeof $F$ with respect to its argument. Denoting by $\vect
x(t)=\big(x(t),y(t),z(t)\big)$ the electron trajectory, the \ALD
equation takes the form
\begin{equation*}
  \left\{
  \begin{aligned}
    m\ddot{x} &=\frac{e}{c}F'(x-ct)\dot{y}+m\veps\dddot{x}
    \\ m\ddot{y} &=
    eF'(x-ct)-\frac{e}{c}\dot{x}F'(x-ct)+m\veps\dddot{y} \\ m\ddot{z}
    &= m\veps\dddot{z} \ ,
\end{aligned}\right.
\end{equation*}
where we have denoted the constant $2e^2/mc^3$ by $\veps$. Notice that
the equation for $z$ decouples, and that the only nonrunaway solutions
are $z(t)=z_0 + v_zt$, i.e., uniform motions. From now on, we
consider just the first two equations, which, by defining $\xi\equal
x-ct$, can be put in the following form
\begin{equation}\label{eq:ALD}
  \left\{
  \begin{aligned}
    \ddot{\xi} &= \frac{e}{mc} F'(\xi)\dot{y} + \veps\dddot{\xi}
    \\ \ddot{y} &= -\frac{e}{mc} F'(\xi)\dot{\xi} + \veps\dddot{y} \ .
  \end{aligned}\right.
\end{equation}

The phase space corresponding to such an equation is six--dimensional,
but the system can be reduced to a four dimensional one exploiting the
invariance by translation along the $y$--axis.  In fact, the second
equation gives
\begin{equation*}
  \frac {\mathrm{d}~}{\mathrm{d}t}\bigg( \dot y - \frac{e}{mc} F(\xi)
  - \veps\ddot{y} \bigg)=0 \ ,
\end{equation*}
i.e., the \ALD equation reduces to
\begin{equation*}
  \left\{
  \begin{aligned}
    \ddot{\xi} &= \frac{e}{mc} F'(\xi)\dot{y} + \veps\dddot{\xi}
    \\
    \dot{y} &= -\frac{e}{mc} F(\xi) + \veps\ddot{y} + C \ ,
\end{aligned}\right.
\end{equation*}
where $C$ is an integration constant,  depending on the initial
data. We can include the constant $C$ in the potential, thus defining the
"effective potential'' $F_C(\xi)=F(\xi)+C$,
and introduce the new variable $v\equal\dot y$: in such a way, one gets
the equation
\begin{equation}\label{eq:1}
  \left\{
  \begin{aligned}
    \dddot{\xi} &= \frac 1\veps\bigg( \ddot \xi -\frac{e}{mc}
    F_C'(\xi)v \bigg)
    \\
    \dot{v} &= \frac 1\veps \bigg( v + \frac{e}{mc} F_C(\xi) \bigg) \ ,
\end{aligned}\right.
\end{equation}
i.e., an equation in a four--dimensional phase space.

To discuss the solution of this equation, consider first the
``mechanical case'' $\veps=0$, i.e., the case in which emission is
neglected. So  one gets
\begin{equation*}
  \left\{
  \begin{aligned}
    \ddot{\xi} &= \frac{e}{mc} F_C'(\xi)v \\ v &= -\frac{e}{mc}
    F_C(\xi) \ ,
\end{aligned}\right.
\end{equation*}
which reduces to the one dimensional Newton's equation
\begin{equation*}
  \ddot{\xi} = -\frac{e^2}{m^2c^2} F_C'(\xi)F_C(\xi) \ ,
\end{equation*}
with a potential $V_C(\xi)= e^2F^2_C(\xi)/2m^2c^2$. The
solutions are readily found. In particular, for motions of scattering
type, if the initial ``kinetic energy'' $\dot \xi^2/2$ is larger then
the maximum of $V_C(\xi)$, the electron will pass the barrier, while
it will be reflected if the initial kinetic energy will be smaller.
In addition, it is easily checked that the zero of $F_C(\xi)$
gives stable equilibrium points, while maxima of the modulus
$|F_C(\xi)|$ will give unstable equilibria.

Return now to the full \ALD equation (\ref{eq:1}). As recalled in the
introduction, we look for for ``exceptional'' initial data which
correspond to solutions having an  asymptotically  vanishing  acceleration.
In other terms, given the initial value $\xi_0$ and
$\dot{\xi_0}$, we want to find whether  initial data $v_0$ and
$\ddot{\xi}_0$ exist such that the corresponding solutions of (\ref{eq:1}) are
nonrunaway. Since we are  considering  a scattering problem, this
can be implemented in a straightforward way by numerically integrating
backward in time the equations of motion. In other terms, one fixes the
final data outside the interaction zone and integrates backwards in
time: in such a way the Dirac manifold\footnote{The Dirac manifold is
 defined as the subset of phase space spanned by the non runaway solutions.}
becomes an actractor, and after a small transient the orbit practically
will lie on such a manifold. Once the electron did come back into the non
interacting zone, one gets the initial data which gives rise to a
nonrunaway solution.

Numerical evidence suggests that, if the electromagnetic field $F(\xi)$
is ``strong'' enough, then, having fixed the mechanical data $\xi_0$
and $\dot{\xi_0}$, there exist several initial $v_0$ and $\ddot{\xi}_0$
which give rise to nonrunaway different trajectories.\footnote{In
  geometric terms, the Dirac manifold is folded.}.  Such nonuniqueness
phenomenon will be discussed in the next Section.

\section{The nonuniqueness phenomenon}\label{sez:nonuni}

Following ref.~\cite{carati95}, in order to discover whether there
exist several nonrunaway solutions corresponding to the same initial
mechanical state, we start investigating the unstable equilibrium
point. By rescaling time by $t\rightarrow \veps t$, the equations
(\ref{eq:1}) becomes
\begin{equation}\label{eq:2}
  \left\{
  \begin{aligned}
    \dddot{\xi} &= \frac{e\veps^2}{mc} F_C'(\xi)v + \ddot{\xi}
    \\ \dot{v} &= -\frac{e}{mc} F_C(\xi) + v \ .
  \end{aligned}\right.
\end{equation}

The equilibrium points of such an equation can be subdivided into two
classes:
\begin{itemize}
   \item The point(s) $v=0,\xi=\xi^*$ with $F_C(\xi^*)=0$ (and obviously
     $\dot \xi=\ddot \xi =0$). Such points corresponds to the stable
     equilibrium points of the mechanical case, and are not
     interesting for the scattering states.\footnote{The nonrunaway
       solutions are the ones which fall on the equilibrium point,
       and thus they do not describe scattering states.}
   \item Points $v=v^*,\xi=\xi^*$ with $F_C'(\xi^*)=0$ and
     $v^*=\frac{e\veps}{mc} F_C(\xi^*)$.
\end{itemize}

We consider only equilibrium points of the second type, more precisely
we considers points such that $\xi^*$ is a maximum for $F_C^2(\xi)$.
It turns out that, as the parameters are changed, such points exhibit a
bifurcation from a saddle to a saddle-focus, the same which drives the
nonuniqueness phenomenon in the one dimensional case (see ref.~\cite{carati95}).
In fact, putting $\chi=\xi-\xi^*$ and $u=v-v^*$, the
equation (\ref{eq:2}) to the first order becomes
\begin{equation}\label{eq:3}
  \left\{
  \begin{aligned}
    \dddot{\chi} &= -k^2\chi + \ddot{\chi} \\
    \dot{u} &= u \ ,
  \end{aligned}\right.
\end{equation}
were we defined
\begin{equation}\label{eq:4}
  k^2 \equal \frac{e^2\veps^2}{m^2c^2} F_C''(\xi^*)F_C(\xi^*) \ .
\end{equation}

So the linearized equations decouple: the second
one defines a direction which is always unstable, while the first one
is the same one just studied for the one-dimensional case in
ref.~\cite{carati95}. As shown in the quoted paper, there is a bifurcation
value
\begin{equation}\label{eq:5}
k_{cr} = \frac {2\sqrt3}{9} \ . 
\end{equation}
For $k<k_{cr}$ the equilibrium point is an unstable saddle, with one
stable direction and two unstable ones. Instead, for $k>k_{cr}$ one
gets two complex eigenvectors, i.e., one has again a stable
one-dimensional manifold
(call it $\Sigma^s$), but the unstable manifold is indeed an unstable
focus: the points spiral out from the origin going to infinity. This
is the source of the nonuniqueness behavior.

In fact, one can argue as follows. Return to the nonlinear 
equation: the unstable manifold is three--dimensional, while the
nonrunaway manifold, as recalled above, is a two-dimensional one, so
that generically there will be a one-dimensional intersection
$\gamma(t)$, which will be a solution belonging both to the unstable
manifold and to the nonrunaway manifold: $\gamma(t)$ springs out
spiraling from the unstable equilibrium point at $t=-\infty$, and goes
to infinity with a vanishing acceleration for $t\to+\infty$.

Consider now, at $t=+\infty$, the nonrunaway solutions near to
$\gamma$, and propagates them back in time: by continuity of solution
of (\ref{eq:2}) with respect to the initial data, this solutions will
follow $\gamma(t)$ near the equilibrium point spiraling about the
stable one-dimensional manifold $\Sigma^s$. The backward-time flow
turns the stable direction into the only unstable one, so that the orbits
will finally follow the $\Sigma^s$ manifold returning again to
infinity. In other terms, the existence of an intersection between the
unstable manifold and the Dirac one, entails that the Dirac manifold
will be wrapped around the stable manifold $\Sigma^s$. This is the
origin of the non uniqueness property.

In fact, fix now $\xi=const$ sufficiently distant from the origin, and
consider the intersection of the two-dimensional Dirac manifold
(before scattering) with the three-dimensional hyperplane
$\xi=const$: one would get a curve which projects on the plane of the
initial ``mechanical data'' $(\dot \xi,v)$ like a (deformed)
spiral. Letting $C$ changing, the different spiral will in general
intersect\footnote{Because in general they will have different
  center.} giving rise to different nonrunaway trajectory for the
same mechanical initial data.

These geometric considerations are obviously not a proof, but just an
indication that the bifurcation of the unstable equilibrium points
could drive the appearance of the nonuniqueness behavior. In the next
Section we will show, by numerical computations, that this is indeed the
case.
\begin{figure}
  \begin{center}
    \includegraphics{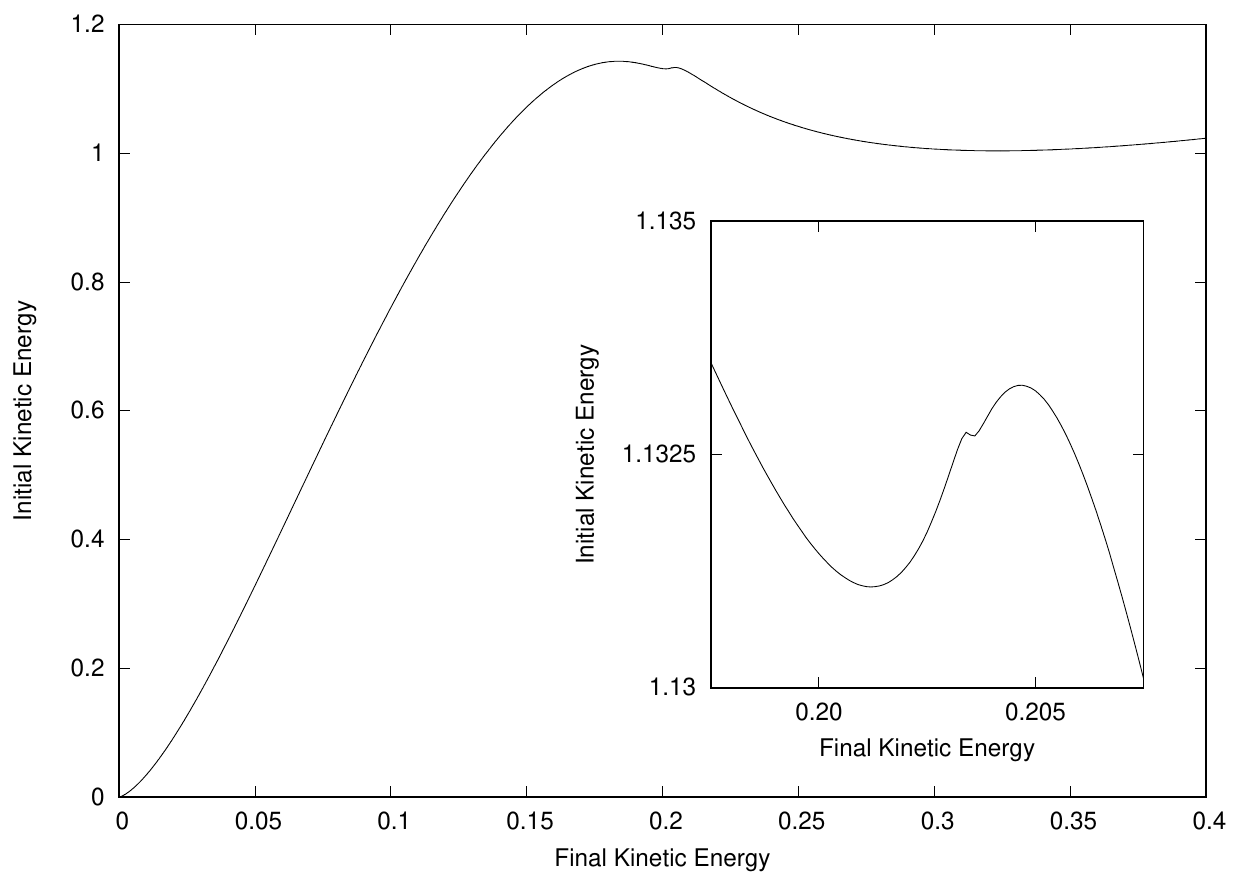}
  \end{center}
  \caption{\label{fig:1} Plot of the initial kinetic energy vs. the
    final one for field amplitude $A=1$, and vanishing wave
    vector. The map is not one to one, and this implies nonuniqueness
    of the nonrunaway solutions. Indeed, drawing a horizontal line at
    energy about $1.13$, one immediately checks that to a given
    initial energy there correspond different final ones. The inset
    hints at the complex structure of the maxima and minima of such a
    curve.}
\end{figure}

\section{Numerical results}\label{sez:numeric}

The equations of motion (\ref{eq:ALD}) were integrated by a third
order Runge--Kutta method which is easy to implement and sufficiently
fast for our purposes.  Moreover, we studied two case: either a simple
Gaussian incoming wave
\begin{equation}\label{eq:6}
   \frac{e\veps}{c} F(\xi) = A e^{-\frac {\xi^2}{\sigma^2}} \ ,
\end{equation}
or the more complex wave form
\begin{equation}\label{eq:7}
   \frac{e\veps}{c} F(\xi) = A e^{-\frac {\xi^2}{\sigma^2}} \;\cos k\xi
   \ ,
\end{equation}
which allows one to investigate the role of the wave-length in the
scattering process. In the latter case one can rescale the distances
by the wave length of the incoming laser pulse.  Then, all the
constants of the problem are resumed into only two parameters: the field
intensity $A$ and the width $\sigma$ of the electromagnetic pulse.
\begin{figure}
  \begin{center}
    \includegraphics{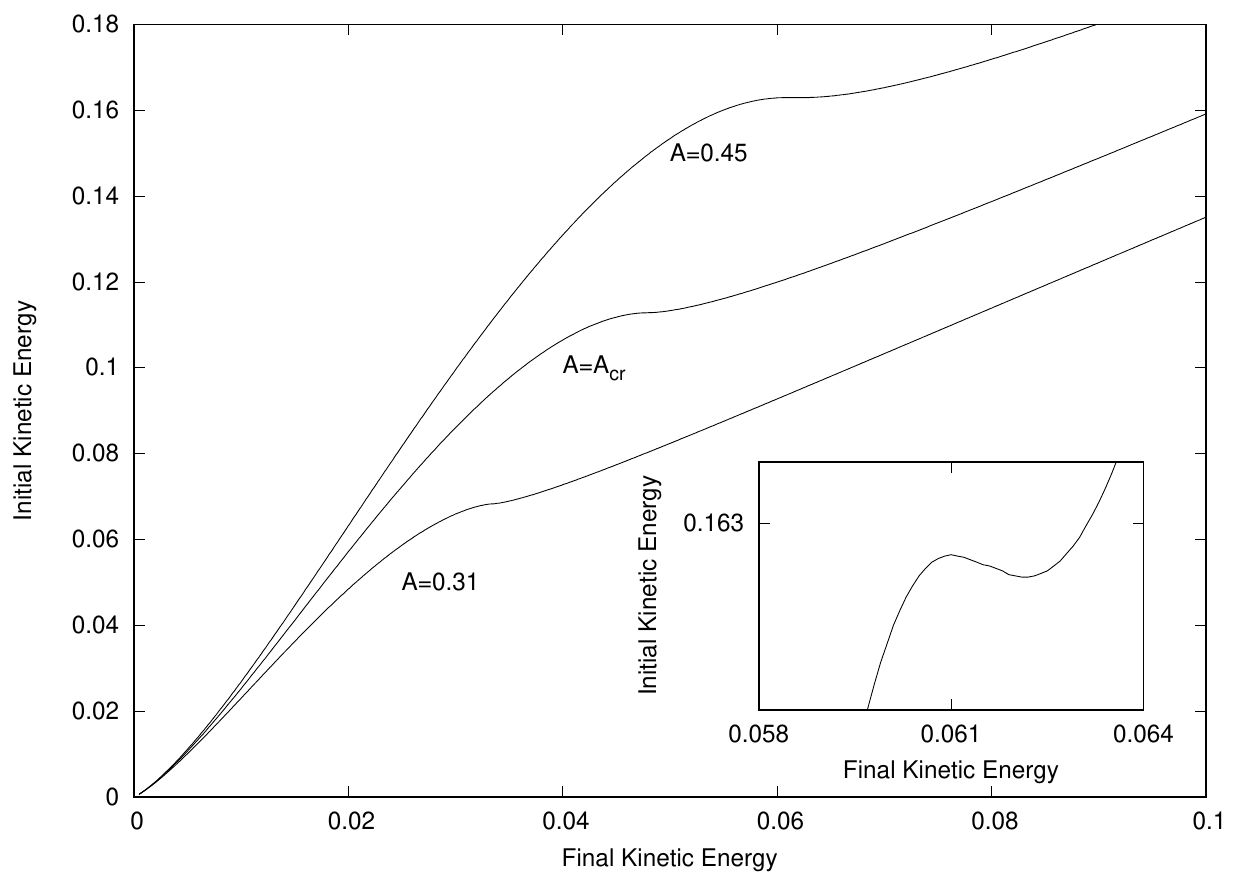}
  \end{center}
  \caption{\label{fig:2} Plot of the initial kinetic energy vs. the
    final one for three field amplitudes: $A=0.31$ smaller than
    $A_{cr}\simeq 0.38$, the critical one and $A=0.45$ larger then the
    critical one.  The wave vector vanishes. One sees that for
    $A=0.45$ there exists a very weak local maximum (see the inset),
    which entails the nonuniqueness of the nonrunaway solutions.}
\end{figure}

In the pure Gaussian case (\ref{eq:6}), we have taken $\sigma =1$ and
studied the behavior of the nonrunaway solutions as the field
intensity $A$ is changed. In particular, we find that $\xi=0$ is an
unstable equilibrium point, in fact the only equilibrium point.  We
compute the value $A_{cr}$ which corresponds, through the formula
(\ref{eq:4}) to the value of $k_{cr}$. For $\sigma=1$ one finds
$A_{cr}\simeq 0.38$.

In the case of the potential given by (\ref{eq:7}), we have taken a
larger value $\sigma =10$, and, by rescaling, $k=1$. In this case
$\xi=0$ is again an unstable equilibrium point, even if now there
exists an infinite number of them (both stable and unstable). The
point $\xi=0$ gives however a lower value for $A_{cr}$, which in
this case corresponds to $A_{cr}\simeq 0.36$.

As remarked in Section~2, to obtain the nonrunaway solutions one
integrate backwards in time.  In such a way, one constructs a map from
the ``final data'' $( \xi^f, y^f,\dot \xi^f, \dot y^f, \ddot
\xi^f,\ddot y^f)$ to the initial one $( \xi^i, y^i,\dot {\xi}^i, \dot
y^i,\ddot {\xi}^i, \ddot y^i)$.  The only independent final parameters
are the final velocities $\dot {\xi}^f$ and $\dot y^f$.  In fact, as
one is dealing with s scattering case, one has to considered $\xi^f$
large (i.e. states in which the electron has left the interaction zone
with the laser pulse), i.e., an arbitrary (but fixed) value for
$|\xi^f|=R$ such that the force due to the electromagnetic field
essentially vanishes.  In such a case one is forced to fix $\ddot
{\xi}^f=\ddot y^f=0$, by the nonrunaway condition. Moreover, due to
the invariance under translation along the $y$ axes, one can fix
arbitrarily $y^f=0$.
\begin{figure}
  \begin{center}
    \includegraphics{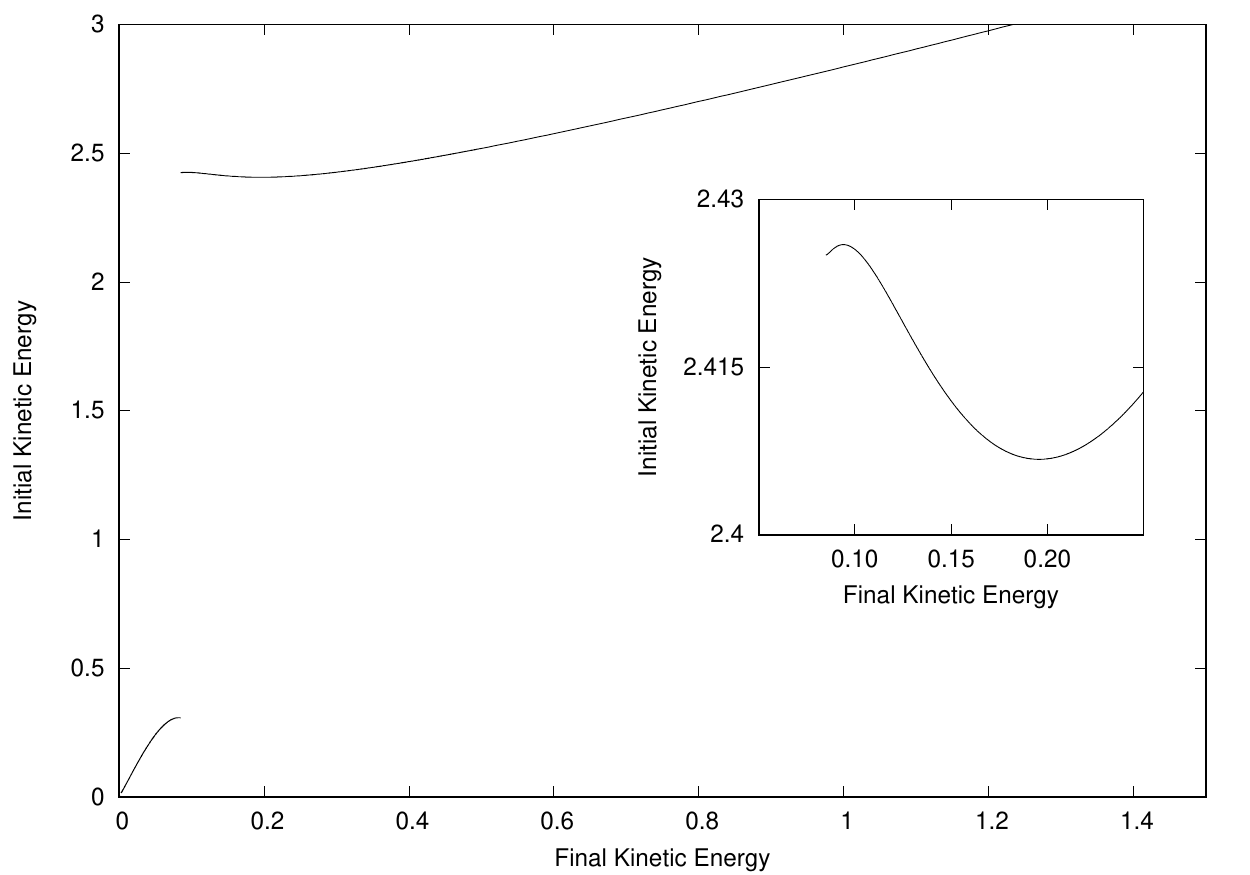}
  \end{center}
  \caption{\label{fig:3} Plot of the initial kinetic energy vs. the
    final one for field amplitude $A=1$, and non vanishing wave
    vector.  Notice the jump. The inset is an enlargement of the
    curves around the minimum, which clearly exhibits the nonuniqueness
    phenomenon. There are other jumps, not shown  in the figure,
    at low energy. The  jumps imply that, for some initial energies,
    there are no scattering solutions: for such energies the electron
    falls onto a stable equilibrium point.}
\end{figure}

Having fixed the final data, one starts  integrating backwards up to
a time such that the electron, after having interacted with the
electromagnetic wave, returns into a zone of vanidhing field,
for example again at $|\xi^i|=R$. At this moment one collects the
initial value $\dot {\xi}^i, \dot y^i, \ddot {\xi}^i, \ddot y^i$. So
defined, the map from the ``final'' to the ``initial'' data is one to
one. The problem is whether the inverse mape, i.e., the physical one which
maps the ``initial'' data to the ``final'' ones, is one to one, or
not. If it is one to one there is uniqueness, i.e., to a mechanical
data of position and velocity corresponds just one nonrunaway
solution; if it is one to many, to a single mechanical state, there
corresponds different nonrunaway solutions with different asymptotic
final states.
 \begin{figure}
  \begin{center}
    \includegraphics{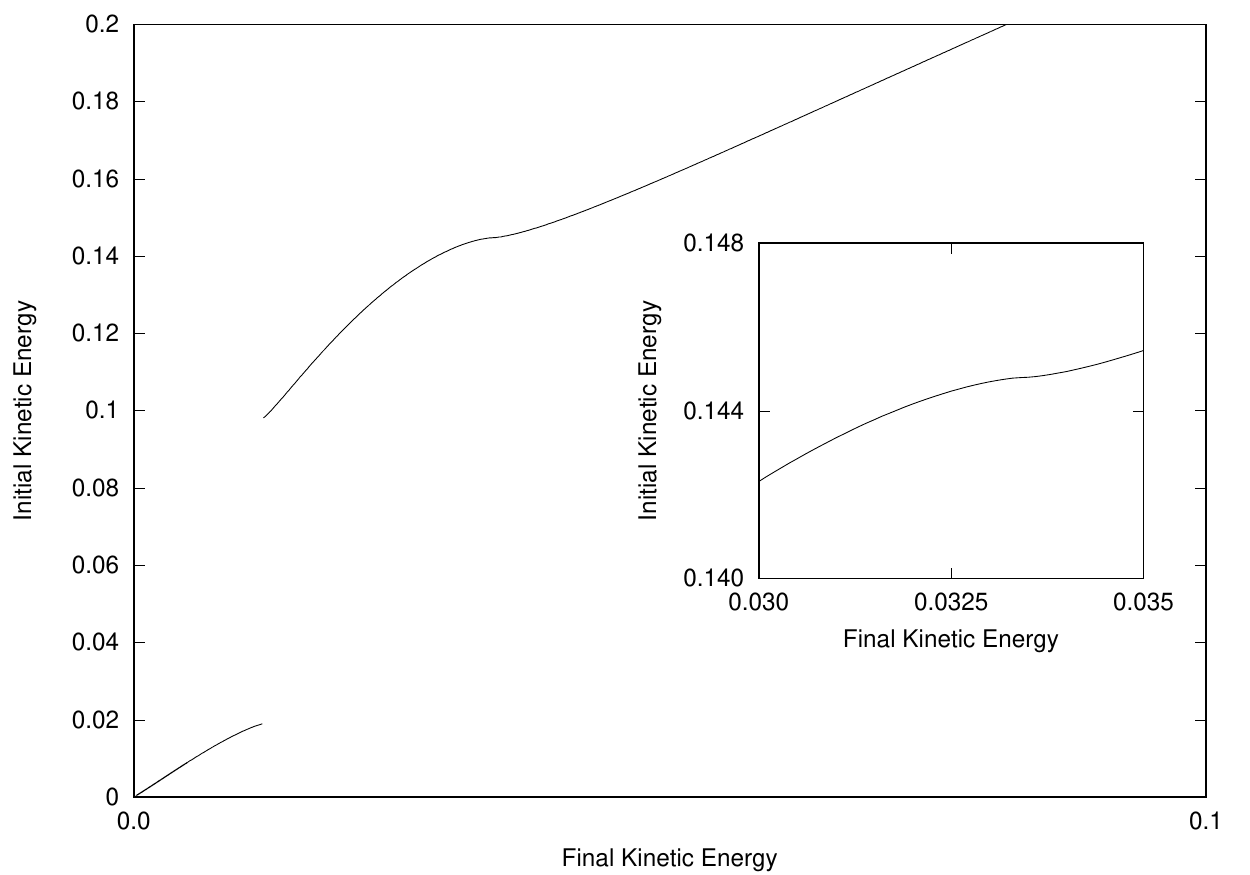}
  \end{center}
  \caption{\label{fig:4} Same as figure~\ref{fig:3}, for $A=0.3$,
    below the critical value. Now the insets show that the map is one
    to one, so that one has uniqueness. }
\end{figure}

In order to answer  this question, we made the preliminary step of reducing
to the case of a scattering normal to the plane wave, i.e., to the case
in which the component $\dot y^i$ along the $y$ axis of the initial
velocity vanishes. So, one has to solve the equation $\dot y^i(\dot
{\xi}^f, \dot y^f)=0$ (which is easily solved by the bisection
method). This gives $\dot y^f$ as a function of $\dot {\xi}^f$, which
remains the only free parameter.  A curious feature of this equation,
probably linked to the conservation of the $y$ component of
momentum in the mechanical case, is that $\dot y^f=0$ gives a good
approximation to the true solution.

Now, by a simple inspection of the curve $\dot {\xi}^i$ as a function of
$\dot {\xi}^f$ one can check whether  the inverse map is one to one or
not. Equivalently one can inspect the curves of the initial kinetic
energy vs. final kinetic energy: in the  nonuniqueness case such
map would show a non monotone behavior. In figure~\ref{fig:1} this
curve is drawn for the pure Gaussian potential (\ref{eq:6}) with
$A=1$, the inset showing details about the local maximum. There is
evidence of a complex sequence of nested maxima, as in the case
investigated in ref.~\cite{carati95}. So the map is not monotone,
and thus the inverse map is not one to one.
\begin{figure}
  \begin{center}
    \includegraphics[width=\textwidth]{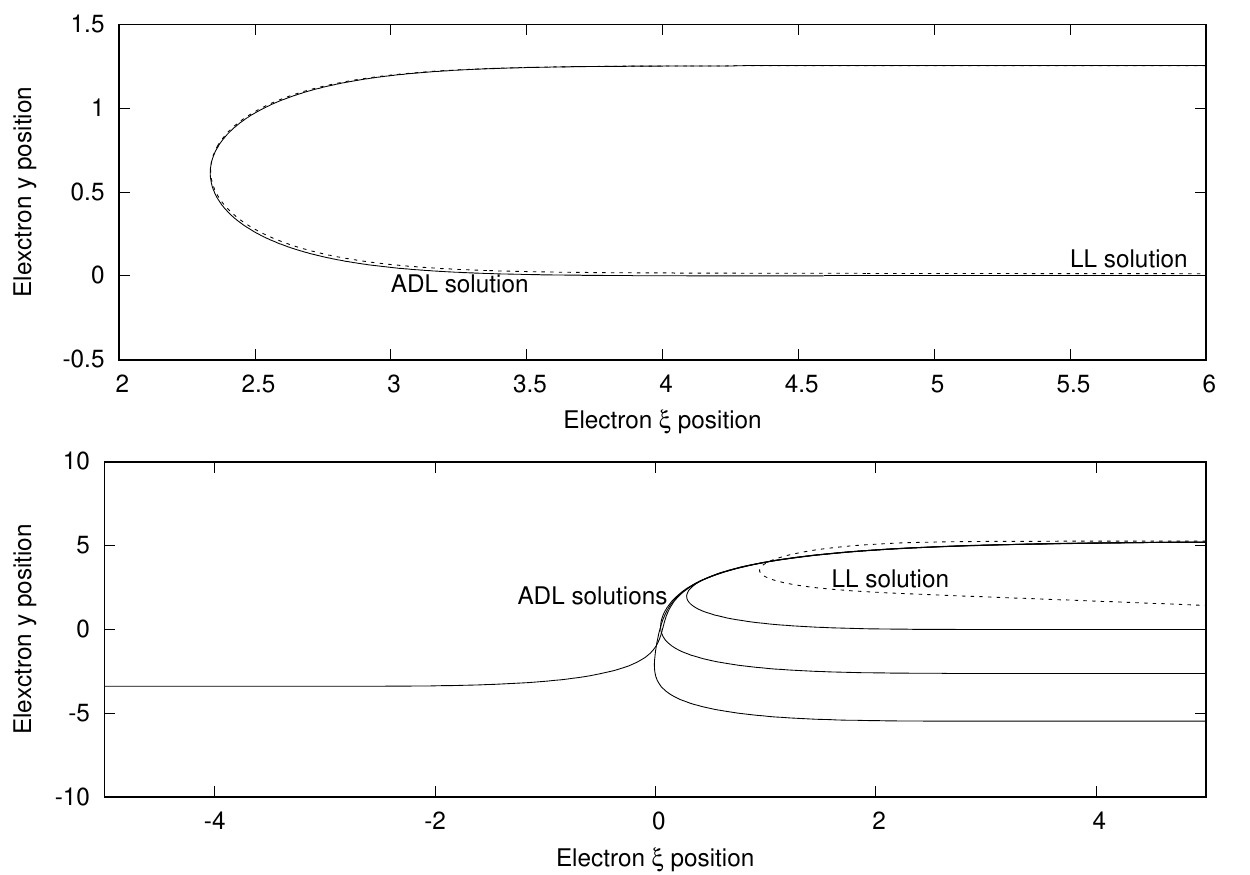}
  \end{center}
  \caption{\label{fig:6} Comparison between orbits computed using the
    Landau--Lifschitz approximation (broken line), and the
    ones computed using the full \ALD equation (full line) with
    the same initial mechanical data: upper panel refers to a pure
    Gaussian field with intensity $A=0.31$ below the critical one and
    initial energy $E=0.00295$; lower panel refers to a pure Gaussian
    field with $A=1$ above the critical one, for an initial energy
    $E=1.1322$ for which there are several  nonrunaway solutions.}
\end{figure}

Figure~\ref{fig:2} shows what happens when the field strength $A$ is
increased from below the critical value to above it:  the curves of the
initial kinetic energy are reported versus the final ones
(always for fixed  initial value $\dot y^i=0$): if $A=0.31$ the curve is
monotone increasing, so that the inverse map is one to one and there
is uniqueness of the nonrunaway solutions. For $A=A_{cr}$ the curve
seems to have an inflection point but one can consider the
inverse map again as one to one, i.e., uniqueness of non runaway
solutions. Instead, a carefully inspection of the case $A=0.41$, above
the critical value, shows a weak local minimum for a final kinetic
energy of $\simeq 0.06$ (more evident in the inset of the figure), so that
uniqueness is lost.

Figure~\ref{fig:3} and \ref{fig:4} refers to the case of the potential
given by (\ref{eq:7}). In figure~\ref{fig:3} the initial kinetic energy  is
plotted versus the final one for $A=1$: the inset is an enlargement
about the minimum. Again, above the critical value, the map is not
one to one, and one has the nonuniqueness phenomenon.  Notice that
the map has a jump, i.e. the inverse map is not defined in a certain
interval. It seem reasonable to assume that for such a  value of the
initial kinetic energy, the incoming particle falls onto one of the
stable equilibrium points. This indeed happens for some value of the energy,
but an analytical proof is lacking.  A more detailed study at low final
energies (too low to be appreciable in the figure) shows that there
are other jumps.

Instead, in figure~\ref{fig:4}, the initial kinetic energy is plotted versus
the final one for $A=0.3$, which is below the critical value. Now, the
map appears to be one to one, and  uniqueness recovered. As in the
case of $A=1$, the inverse map is not defined for some intervals of
the initial kinetic energy. Again we think this is due to the fact
that the particle be captured by one of the stable equilibrium points.
\begin{figure}
  \begin{center}
    \includegraphics[width=\textwidth]{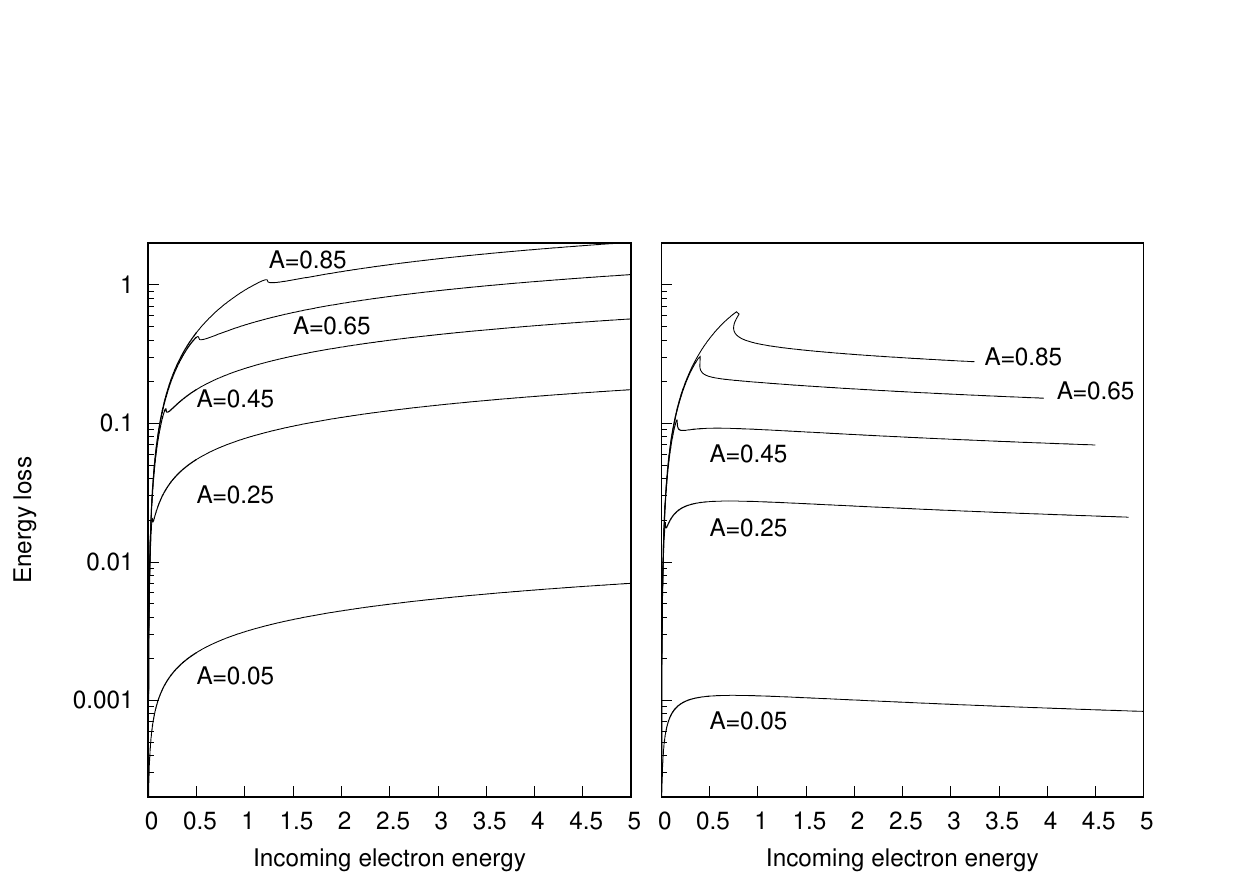}
  \end{center}
  \caption{\label{fig:7} Energy loss in a collision versus the energy
    of the incoming electron, for several values of the field
    intensity, ranging from $A=0.05$, well below the critical
    intensity, to $A=0.85$, above it, in semi logarithmic scale. Left
    panel: loss computed according Landau-Lifshitz approximation;
    right panel: loss computed according \ALD equation. Notice taht the \ALD
    equation predicts a maximum for the energy loss. }
\end{figure}

\section{Comparison with the Landau--Lifschitz approximation}\label{sez:landau}

The Landau-Lifschitz approximation is obtained from the \ALD equation
using the following argument. For small $\veps$ one has
\begin{equation}\label{eq:8}
  m \ddot{\vect x} \simeq \vect F(\vect{x}, \dot{\vect x}) \ ;
\end{equation}
so that one can obtain an approximation of the third derivatives by
\begin{equation*}
  \dddot{\vect x }= \frac {\diff~}{\diff t} \ddot{ \vect x } \simeq
  \frac {\diff~}{\diff t} \Bigg( \frac 1m \vect F(\vect x, \dot{\vect
    x})\Bigg) \ ,
\end{equation*}
which substituted into the \ALD equation gives, neglecting the terms of
order higher, 
\begin{equation}
  m \ddot{\vect x} = \vect F(\vect{x}, \dot{\vect x}) + \veps \Big(
  \frac{\partial \vect F}{\partial \vect x}\dot{\vect x} + \frac 1m
  \frac{\partial \vect F}{\partial \dot {\vect x}}\vect F \Big)\ ,
\end{equation}
where we have replaced $\ddot {\vect x}$ again by its approximation
(\ref{eq:8}).  This equation does not have the problem of runaways and
therefore of the choice of initial data.

Using a third order Runge--Kutta methods, we integrate this equation,
with the Gaussian vector potential as given by (\ref{eq:6}), for
several values of the intensity $A$. Figure~\ref{fig:6} show the
orbits found: they are computed by first integrating the \ALD equation
backward for a certain amount of time, and then the Landau--Lifschitz
equation forward in time, so that the initial mechanical data for the
two equations agree. One can check that the orbits for low values of
$A$ essentially coincide, while they differ for higher field
intensities as expected.

A more meaningful comparison is given in figure~\ref{fig:7}, where is
reported the loss of energy in a collision versus the energy of the
incoming electron, for different values of the field intensity,
ranging from $A=0.05$ to $A=0.85$: in the left panel the loss is
computed from the Landau--Lifschitz, while on the right the loss is
computed according the \ALD equation. They are qualitatively different
also for field's intensities well below the threshold, inasmuch as
according \ALD the loss has a well defined maximum at a definite
energy, while according Landau--Lifschitz it keeps increasing
without limit, at least in the range of energy we have explored. In
addition, the energy loss is systematically smaller for the \ALD
equation with respect to its approximation.

\section{Conclusion}\label{sez:concl}

We have show that, for the nonrelativistic \ALD equation, there
exists a threshold for the intensity of the incoming field, above
which one has several  nonrunaway solutions for the same mechanical
initial data of position and velocity. Such a threshold agrees well
with the bifurcation value of the main unstable point from saddle to
saddle--focus.

Above such a threshold the Landau-Lifschitz approximation is clearly
in defect, but we have checked that also below threshold the loss of
energy in a collision does not agree. As this is the main experimental
observable, one may wonder whether the use of the full \ALD equations
instead of the Landau--Lifschitz approximation might lead to an
agreement between theory and experiment, better than the one found in
ref.~\cite{dipiazza}.

Answering  such a question  would require an analysis similar to that
performed in this pape,  for the full relativistic \ALD equation, in the
regime of interest for the experiments. This is a much more complex task,
on which  we hope to come back  in the future.

\vspace{2.em}
\noindent
\textbf{Acknowledgments}. We thank prof.~Galgani for indicating to us
the role the full \ALD equation might present in explaining the
experiments of scattering of electrons by strong laser pulses, and for
many discussions and suggestions.

\end{document}